\documentclass[10pt,aps,twocolumn,showpacs,preprintnumbers,amsmath,amssymb,prl]{revtex4-2}
\usepackage{graphicx}
\usepackage{dcolumn}
\usepackage{bm}
\usepackage{amsmath}
\usepackage{color}
\usepackage{lineno}

\usepackage{booktabs}

\begin{document}
\draft
\title{Activity-assisted  barrier-crossing of self-propelled colloids over parallel microgrooves}
\author{Yan Wen,$^{1}$ Zhihao Li,$^{2}$ Haiqin Wang,$^{2,3}$ Jing Zheng,$^{4}$ Jinyao Tang,$^{4}$ Pik-Yin Lai,$^{5,6,\ast}$ Xinpeng Xu,$^{2,\ast}$ Penger Tong$^{1,}$}\email{Corresponding emails: pylai@phy.ncu.edu.tw, xu.xinpeng@gtiit.edu.cn, and penger@ust.hk}
\affiliation{[1] Department of Physics, Hong Kong University of Science and Technology, Clear Water Bay, Kowloon, Hong Kong\\
[2] Physics Program, Guangdong Technion-Israel Institute of Technology, Shantou, 515063, China\\
[3] Technion-Israel Institute of Technology, Haifa, 3200003, Israel\\
[4] Department of Chemistry, The University of Hong Kong, Pokfulam, Hong Kong\\
[5] Department of Physics and Center for Complex Systems, National Central University, Taoyuan City 320, Taiwan\\
[6] Physics Division, National Center for Theoretical Sciences, Taipei 10617, Taiwan}
\date{\today}
\begin{abstract}
We report a systematic study of the dynamics of self-propelled particles (SPPs) over a one-dimensional periodic potential landscape $U_0(x)$, which is
fabricated on a microgroove-patterned polydimethylsiloxane (PDMS) substrate. From the measured non-equilibrium probability density function $P(x;F_0)$ of the SPPs, we find that the escape dynamics of the slow rotating SPPs across the potential landscape can be described by an effective potential $U_{\text{eff}}(x;F_0)$, once the self-propulsion force $F_0$ is included into the potential under the fixed angle approximation. This work demonstrates that the parallel microgrooves provide a versatile platform for a quantitative understanding of the interplay among the self-propulsion force $F_0$, spatial confinement by $U_0(x)$ and thermal noise, as well as its effects on activity-assisted escape dynamics and transport of the SPPs.
\end{abstract}
\pacs{}
\maketitle

Self-propelled colloids are synthetic micro-swimmers that can take up energy from the surrounding environment and convert it into directed motion \cite{Poon13,Bechinger16,Zhang2017,Wang20}. They are developed as simplified model systems that feature similar dynamic behaviors of various living systems, such as molecular motors, cells and bacteria \cite{Ross08,Zajac13,Berg2004,Cates2012}. The micro/nanoscale swimmers hold great promise for a wide range of biomedical applications, ranging from directed cargo transport and drug delivery to targeted cancer therapy \cite{Zhang2017,Brambilla2001,Wang2012}. These applications require a better understanding of the dynamics of individual self-propelled particles (SPPs) and their response to complex environments, which can be modeled by imposing well-characterized external potential (or free-energy) landscapes. Unlike passive colloids at equilibrium, whose statistical distribution and escape dynamics over a potential landscape can be understood through Boltzmann statistics and Kramers theory \cite{Hanggi90,Evans2001,Ma13}, SPPs are intrinsically in a non-equilibrium state \cite{Ramaswamy2010,Marchetti2013,Marchetti2016,Zottl2016}, in which detailed balance and Boltzmann distribution do not apply in general.

A number of experiments have been carried out to study the dynamics of SPPs under the influence of simple potential fields $U_0(x)$, such as linear gravitational potentials \cite{Ginot18,Palacci10} and two-dimensional (2D) optical or acoustic traps \cite{Takatori16,Shen19,Schmidt21}.
These studies found that depending on the interplay between the external potential force, $F_e(x)\equiv -dU_0(x)/dx$, and the self-propulsion force $F_0$ (which is proportional to the self-propulsion velocity $v_0$ of SPPs), the SPPs can stay in two dynamically distinct states. One is the ``bound state" in which the SPPs spend more time at the force balanced location, resulting in a local accumulation of the SPPs near the confining boundary \cite{Takatori16,Shen19,Schmidt21,Volpe11}. This is achieved when the persistence time $\tau_p$ of the SPPs is much longer than their travelling time between the confining boundaries. The other is the ``extended state" in which either $F_e(x)$ is too small to balance $F_0$ or the lifetime of the bound state is too short, so that the SPPs can explore all the available space of $U_0(x)$ at an elevated effective temperature \cite{Palacci10,Maggi14,Choudhury17}.

In fact, there are many practical situations of interest remaining between the two limiting cases. For example, the motion of SPPs along a thin channel or a groove is a common way for active particle transport \cite{Palacci13}, in which the SPPs are confined only in the lateral direction but are free to move along the longitudinal direction. Because there are fewer experimental systems in which one can actually visualize the potential landscape and track the individual SPPs with sufficient statistics, much of the work done so far in this area is through computer simulations \cite{Pototsky12,Geiseler16,Solon15,Fodor16,Sharma17,Woillez19}. Experiments with well-characterized SPPs are, therefore, very valuable in testing different ideas and providing new insights into the non-equilibrium statistical properties of active colloids.

In this Letter, we present a combined experimental and theoretical study of the dynamics of a dilute monolayer of SPPs over a substrate with parallel microgrooves that are fabricated by photolithography. With a large volume of the SPP trajectories obtained using optical microscopy and multi-particle tracking, we are able to provide a statistical description of the non-equilibrium distribution of the SPPs over a 1D periodic potential landscape. A central finding of this investigation is that the non-equilibrium behaviors of the SPPs in the limit of long persistence time $\tau_p$ can be described by an effective equilibrium approach, once the self-propulsion force $F_0$ is properly included into the effective potential $U_{\text{eff}}(x;F_0)$.

\begin{figure}
\centering
\includegraphics[width=0.45\textwidth]{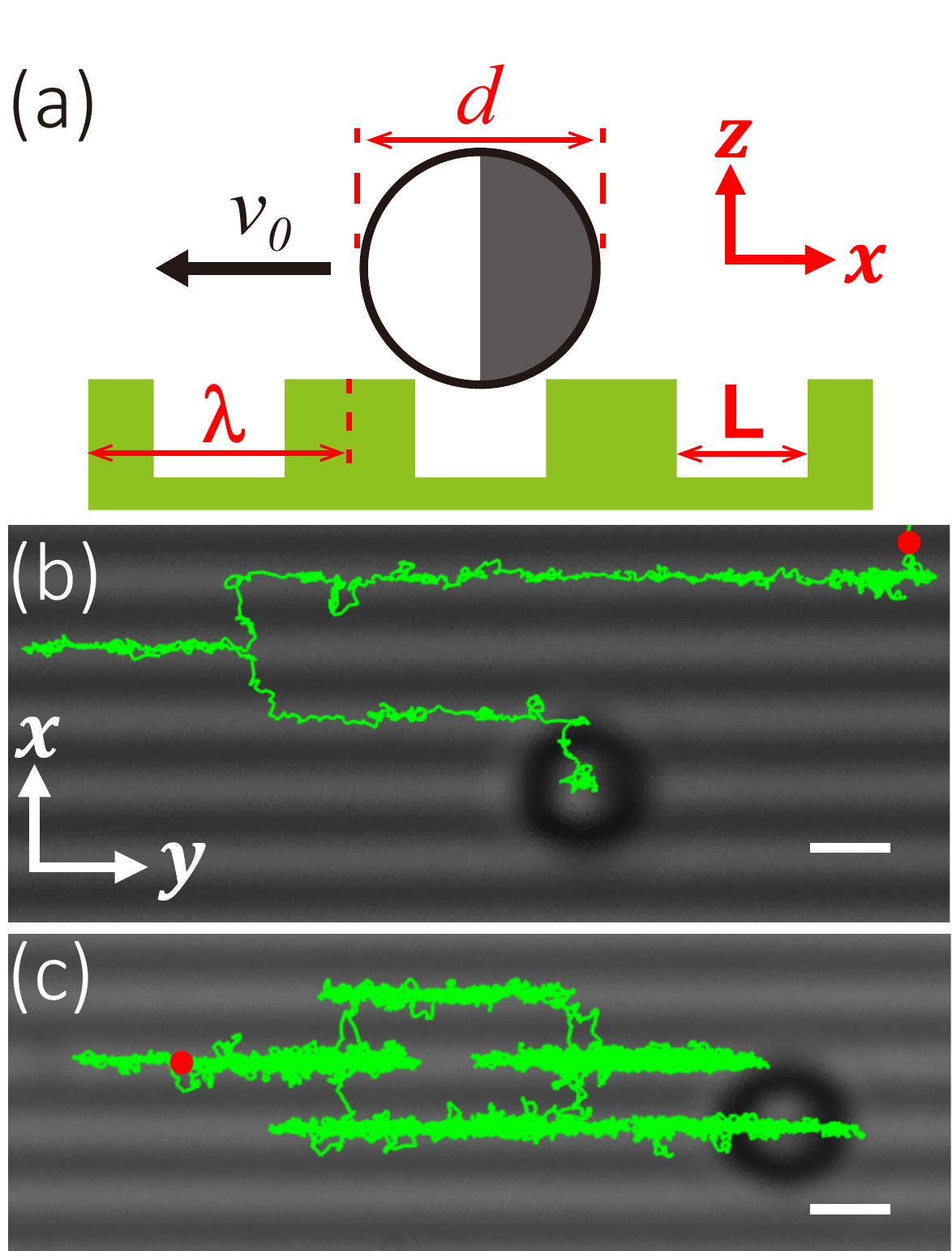}
\caption{(a) A schematic side view of a Pt-SiO$_2$ Janus particle of diameter $d$ moving over a microgroove-patterned substrate with groove spacing $L$ and period $\lambda$. The grey hemisphere indicates the Pt coating and the black arrow indicates the direction of the self-propulsion velocity $v_0$.
(b) \& (c): Two representative trajectories of the Janus particle in the aqueous solution of 0.2\% H$_2$O$_2$ (b, duration $\sim$200 s) and in water (c, duration $\sim$0.5 h), respectively. The data are taken from Sample S1. The parallel stripes in the background show the microgroove pattern with alternating ridges (darker stripes) and grooves (brighter stripes). The dark circle shows the Janus particle image at the beginning of the trajectory.
The red dots indicates the end of the trajectory. Also shown in (a) and (b) are the coordinate systems used in the experiment. The scale bars are 2 $\mu$m.}
\label{Figure01}
\end{figure}

The SPPs chosen for the study are monodisperse SiO$_2$ micro-spheres, whose half surface is coated
with a thin layer of platinum of thickness $\sim$5 nm via sputter coating. The coating is so thin that the Pt-SiO$_2$ Janus particles can rotate freely in 3D with minimal influence from an uneven gravitational torque. When dispersed in an aqueous solution of H$_2$O$_2$, the self-diffusiophoresis of the Janus particles gives rise to a self-propulsion velocity $v_0$, which increases with the H$_2$O$_2$ concentration \cite{Note1,Howse07}. A low H$_2$O$_2$ concentration in the range of 0--1.5 w/w\% is used to avoid O$_2$-bubble formation. This is a well-studied SPP system whose synthesis procedures are known \cite{Zhang2017,Wang20} (see Supplementary Sec.~I.B for more details \cite{Note1}).

The sample cell is a thin circular fluid chamber of diameter 2.2 cm and is made of stainless steel. The bottom coverslip is coated with a thin layer of polydimethylsiloxane (PDMS) substrate, containing a parallel array of microgrooves as illustrated in Fig.~\ref{Figure01}(a). When a dilute monolayer of the Janus particles move over the rugged surface of the microgroove-patterned substrate, they experience a 1D periodic gravitational potential $U_0(x)$ in the $x-$direction normal to the microgroove \cite{Ma13,Su17}. The particle's motion along the $y-$direction parallel to the microgroove is unrestricted. The potential $U_0(x)$ is determined primarily by the ratio of the particle's diameter $d$ to groove spacing $L$ and is not very sensitive to the groove depth, which is fixed at 0.6 $\mu$m, because the Janus particles are suspended.
The area fraction $n_p$ occupied by the Janus particles is less than 0.7\%, at which the interactions between the particles can be ignored \cite{Ma13,Ma15}. In the experiment, we use two samples with different potential fields $U_0(x)$ and their parameters are given in Table~\ref{Table01}. The particle's motion is viewed under bright field microscopy and recorded by a CMOS camera at a sampling rate of 10 frames per second (see Supplementary Sec.~I for other experimental details \cite{Note1}).

Figures~\ref{Figure01}(b) and \ref{Figure01}(c) show two representative trajectories of a Janus particle when it is in the  active state with H$_2$O$_2$ supplied and in the passive state with no H$_2$O$_2$ supplied, respectively. Moving over the parallel microgrooves, the particles do not undergo  unrestricted lateral motion, as was observed over a flat substrate \cite{Howse07}. Instead, their trajectories reveal a confined motion within a single groove with occasional hopping to a neighboring groove. The passive trajectory appears more diffusive compared with the active one.
With this setup, we obtain a large volume of particle trajectories over long durations (60--1200 s) by using a homemade single-particle tracking program at a spatial resolution of $\sim$100 nm.

\begin{table}
\centering
\begin{ruledtabular}
\begin{tabular}{lccccccc}
		Sample&$d$ &$L$ &$\lambda$ &$D_0$  &$\tau_p$ &$t_0$&$E_{b}(0)$\\
		&\footnotesize{($\mu$m)} &\footnotesize{($\mu$m)} &\footnotesize{($\mu$m)} &\footnotesize{($\mu$m$^2$/s)} &\footnotesize{(s)} &\footnotesize{(s)} &\footnotesize{($k_BT$)} \\
		\colrule
		S1&2.96&1&2&0.08&11$\pm$ 3 & 0.18$\pm$0.03 & 5.7$\pm$0.02 \\
		S2&2.4&2&3&0.09&7$\pm$ 2 & 0.28$\pm$0.04 & 7.6$\pm$0.05
\end{tabular}
\end{ruledtabular}
\caption{Two samples used in the experiment with particle diameter $d$, groove spacing $L$, groove period $\lambda$, particle's diffusion coefficient $D_0$, persistence time $\tau_p$, relaxation time $t_0$ and equilibrium energy barrier height $E_{b}(0)$.}
\label{Table01}
\end{table}
\begin{figure*}
\centering
\includegraphics[width=1\textwidth]{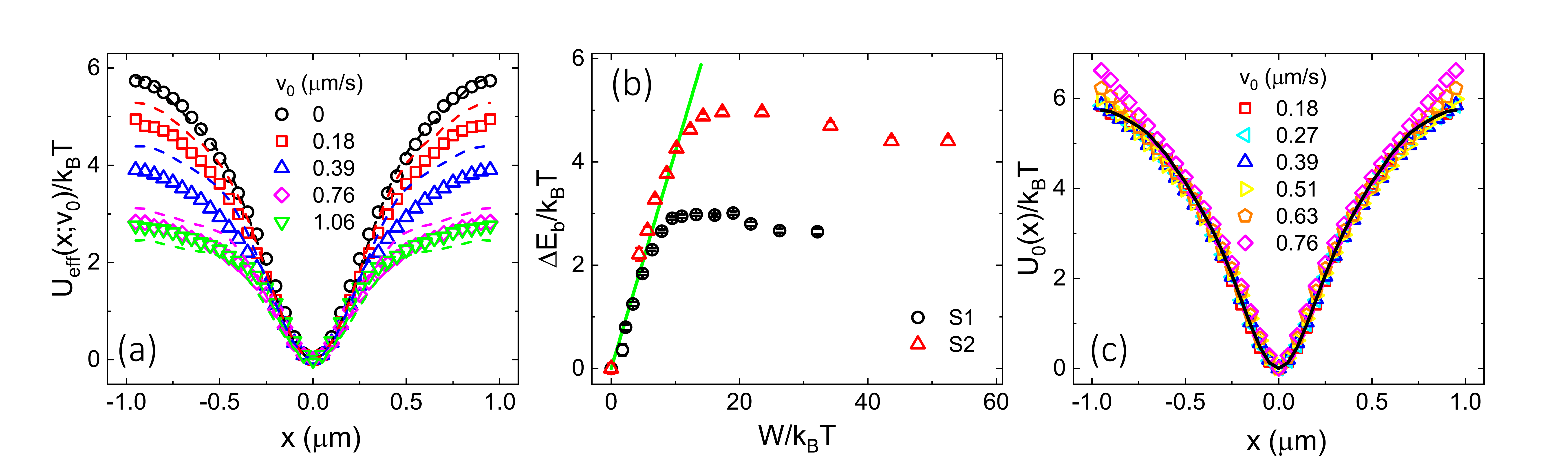}
\caption{(a) Comparison of the effective potentials $U_{\text{eff}}(x;v_0)/k_BT$ that are measured in experiments (open symbols) and calculated in simulation (dashed lines) across one microgroove period. Here $x=0$ is set at the bottom center of the microgroove. The data are taken from Sample S1. The color-coded symbols show the obtained $U_{\text{eff}}(x;v_0)/k_BT$ for increasing values of $v_0$. (b) Measured barrier height difference, $\Delta E_b/k_BT=[E_{b}(0)-E_b(v_0)]/k_BT$, as a function of the normalized self-propulsion work $W/k_B T=v_0\lambda/(2D_0)$
for Samples S1 (black circles) and S2 (red triangles). The solid line shows a linear fit, $\Delta E_b=aW$ with
$a=0.42$, to the data points with small values of $W/k_BT$. The error bars show the experimental uncertainties of the measurements. (c) Reconstructed potential $U_0(x)/k_BT$ from the measured $U_{\text{eff}}(x;v_0)/k_BT$ using Eq.~(\ref{eq02}) for Sample S1 with different values of $v_0$. The black solid line shows the directly measured $U_0(x)/k_BT$ for passive particles with $v_0=0$.}
\label{Figure02}
\end{figure*}

From the obtained particle trajectories, we compute the mean-squared displacement (MSD) $\langle \Delta y^2(\tau)\rangle$ \cite{Howse07},
\begin{equation}
\langle \Delta y^2(\tau)\rangle=2D_0\tau+\frac{2}{3}v_0^2\tau _p\left[\tau+\tau _p(e^{-\tau/\tau _p}-1)\right],
\label{eq01}
\end{equation}
where $\Delta {y}(\tau)={y}(t+\tau)-{y}(t)$ is the longitudinal displacement of the Janus particles along the microgroove with
$D_0$ being their translational diffusion coefficient, which is known from the measured $\langle \Delta y^2(\tau)\rangle$ at $v_0=0$.
With the long particle trajectories, we are able to obtain the trajectory-based velocity $v_{0i}$ from $\langle \Delta y^2(\tau)\rangle_i$ for the ith particle at small values of $\tau$ using Eq.~(\ref{eq01}). We then divide all the labeled particle trajectories with different $v_{0i}$, which are obtained at different H$_2$O$_2$ concentrations, into different subgroups based on a common set of velocity bins. With this procedure, we remove the effect of sample polydispersity in $v_{0}$ resulting from non-uniform coating of the Janus particles (see supplementary Sec.~II.A for more details \cite{Note1}).

With the well-characterized particle trajectories, we obtain the population probability density function (PDF) $P(x,y;v_0)$ of finding a Janus particle with a self-propulsion speed $v_0$ at the location $(x,y)$, where the coordinates are shown in Fig.~\ref{Figure01}(b). By averaging the particle trajectories along the microgroove ($y$-axis) and over different grooves, we obtain the 1D PDF $P(x;v_0)=\langle P(x,y;v_0)\rangle_y$, from which we define the effective potential, $U_{\text{eff}}(x;v_0)/k_BT = -\ln [P(x;v_0)/P(0;v_0)]$, where $P(x;v_0)$ is normalized by the PDF $P(0;v_0)$ at the bottom ($x=0$) of the potential well. In particular, the equilibrium potential, $U_0(x)\equiv U_{\text{eff}}(x;v_0=0)$, is obtained from the passive particles with $v_0=0$.

Figure~\ref{Figure02}(a) shows how the measured $U_{\text{eff}}(x;v_0)/k_BT$ evolves with increasing $v_0$.
The measured equilibrium potential $U_0(x)$ has a symmetric single-well shape with its local minimum centered at the bottom of the microgroove ($x=0$) and its maximum at the middle of the microgroove ridge ($x=\pm \lambda/2$). We now define the effective barrier height $E_{b}(v_0)\equiv U_{\text{eff}}(\lambda/2;v_0)$ and its equilibrium value $E_{b}(0)\equiv E_{b}(v_0=0)$ (which is shown in Table~\ref{Table01} for the two samples). As $v_0$ increases, the measured $U_{\text{eff}}(x;v_0)/k_BT$ remains a similar shape, but its barrier height $E_{b}(v_0)$ decreases with $v_0$ until $v_0 \simeq 0.76~\mu$m/s, above which the measured $U_{\text{eff}}(x;v_0)/k_BT$ does not change with increasing $v_0$ any more and its barrier height saturates at $E_b(v_0)=2.8\pm 0.1~k_BT$. Note that for all the values of $v_0$ studied, no particle accumulation is observed near the confining boundaries (i.e., near the microgroove ridges).

To show how the effective barrier height $E_b(v_0)$ changes with $v_0$, we define the barrier height difference $\Delta E_b=E_{b}(0)-E_b(v_0)$ and normalize $v_0$ as $v_0\lambda/(2D_0)=W/k_B T$, where $W$ is the work done by the self-propulsion force $F_0=v_0/\mu_0$ over the distance of a half microgroove period $\lambda/2$. Here $\mu_0 =D_0/k_BT$ is the mobility of the Janus particles.
As shown in Fig.~\ref{Figure02}(b), the measured $\Delta E_b$ for small $W$ (or for $v_0< 0.76~\mu$m/s before the saturation) can be well described by a linear function, $\Delta E_b=aW$ with the fitting parameter $a=0.42\pm 0.05$.
This result prompts us to consider an effective potential,
\begin{equation}
\widetilde{U}_{\text{eff}}(x;v_0,\theta,\varphi)=U_0(x)-F_0\sin\theta\cos\varphi |x|,
\label{eq010}
\end{equation}
where $F_0\sin\theta\cos\varphi$ is the projection of the self-propulsion force $F_0\hat{\bf{u}}$ along the $x$-axis with $\hat{\bf{u}}$ denoting the orientation of the SPPs in 3D motion, specified by the polar angle $\theta$ and azimuthal angle $\phi$. This activity-modified potential was introduced in Refs.~\cite{Pototsky12,Geiseler16} under the fixed angle approximation, which assumes that the SPPs rotate very slowly with their persistence time $\tau_p$ much longer than their relaxation time $t_0 \equiv k_BT/(kD_0)$ in the potential $U_0(x)$ (with a spring constant $k$). In this limit, the particle's orientation remains unchanged during an escape attempt from the potential (see the values of $t_0$ in Table~\ref{Table01}), and thus $F_0$ will decrease the energy barrier effectively and assist the escape of the SPPs across the 1D potential $U_0(x)$.

In the experiment, $U_{\text{eff}}(x;v_0)$ is obtained by averaging over particle trajectories with all possible orientations. Assuming that the average of the self-propulsion force $F_0\hat{\bf{u}}$ over different orientations $\hat{\bf{u}}$ in 3D can give rise to an effective 1D force $\langle F_0\rangle_{1D}= aF_0$ along the $x$-direction, we have
\begin{equation}
U_{\text{eff}}(x;v_0)\simeq U_0(x)-aF_0|x|,
\label{eq02}
\end{equation}
where the prefactor $a$ ($<1$) is a fitting parameter characterizing the averaging effect of the 3D rotation projected onto the $x$-axis. For a uniformly distribution of $\theta$ and $\varphi$, one may have a simple estimate of $a$ as $a\approx (1/\pi^2)\int_0^\pi \sin(\theta) d \theta \int_{-\pi/2}^{\pi/2} \cos(\varphi) d \varphi=4/\pi^2 \simeq 0.41$, which is close to the experimental value $a=0.42\pm0.05$ (see supplementary Sec.~IV for more discussions \cite{Note1}).

The activity-induced term, $-aF_0|x|$, in Eq.~(\ref{eq02}) measures the difference between $U_{\text{eff}}(x;v_0)$ and $U_0(x)$. Once this term is subtracted out from the measured $U_{\text{eff}}(x;v_0)$, all of the data sets obtained at different $v_0$ for $v_0\leq 0.76~\mu$m/s collapse onto a common master curve, as shown in Fig.~\ref{Figure02}(c). It is seen that the reconstructed data sets agree well with the directly measured $U_0(x)$ from the passive particles (black solid line). Similar results are also obtained for Sample S2, as shown in supplementary Fig.~S7
(see supplementary Sec.~II.C for more details \cite{Note1}). Figures~\ref{Figure02} and S5 thus verify the prediction of Eq.~(\ref{eq02}).
Moreover, from Eq.~(\ref{eq02}), one can find a critical self-propulsion force $F_c\simeq E_{b}(0)/[a(\lambda/2)]$, or a critical velocity $v_c=\mu_0F_c$, above which the effective barrier to escape vanishes and Eq.~(\ref{eq02}) does not hold any more \cite{Evans2001,Ma15}.
For S1, we find $v_c\simeq 1.1~\mu$m/s and hence Eq.~(\ref{eq02}) applies only for $v_0<v_c$. This is consistent with the results shown in Fig.~\ref{Figure02}(c) (and Fig.~\ref{Figure02}(b)) that the scaling of the reconstructed potential works only for $v_0\leq 0.76~\mu$m/s$<v_c$.

\begin{figure}
\centering
\includegraphics[width=0.48\textwidth]{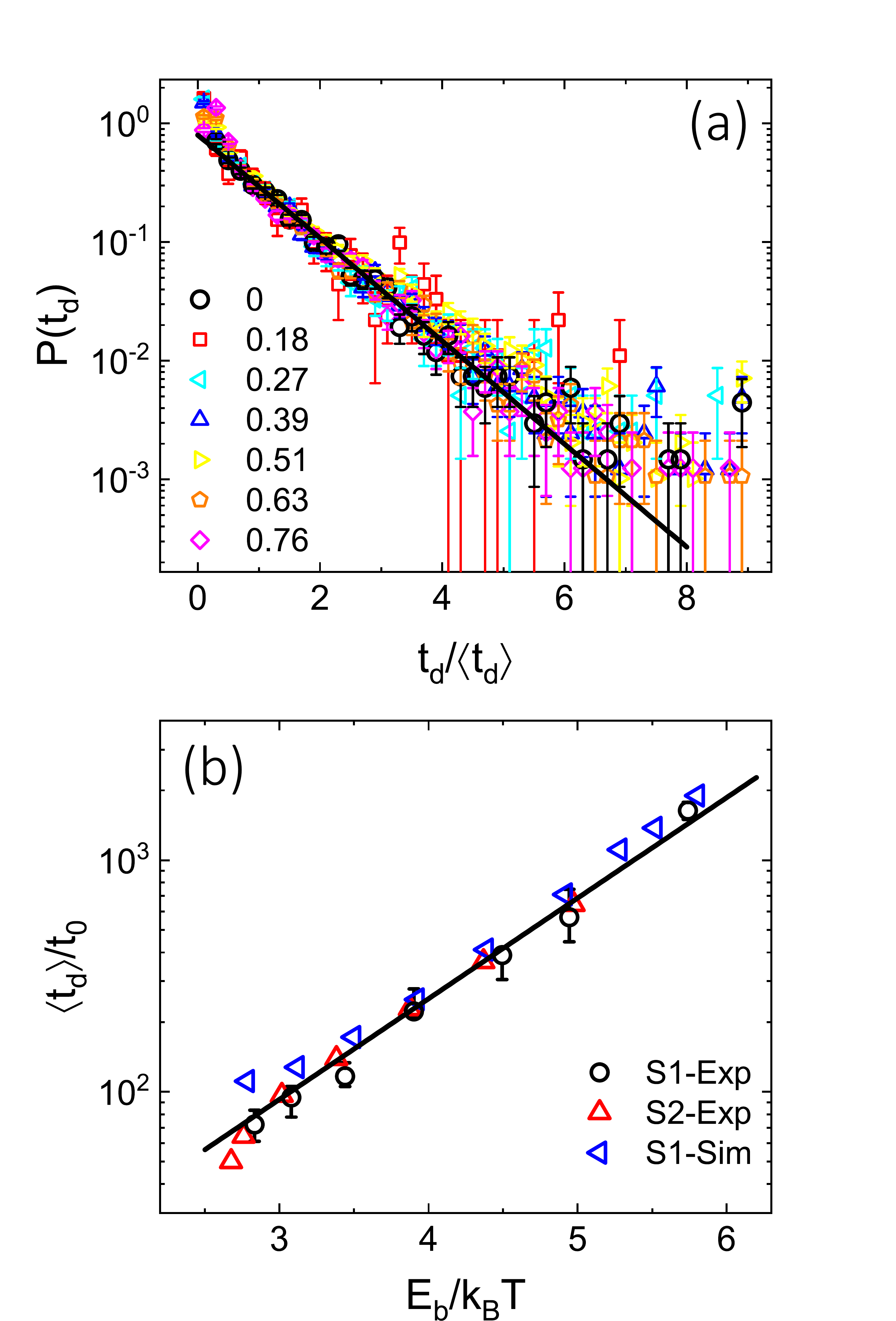}
\caption{(a) Measured PDF $P(t_d)$ of the dwell time $t_d$ for Sample S1. In the plot, $t_d$ is normalized by its mean value $\langle t_d\rangle$ obtained for different values of $v_0$. The solid line shows an exponential fit, $P(t_d)\propto \exp(-t_d/\langle t_d\rangle)$, to the tail part of the measured $P(t_d)$. (b) Experimentally measured (black and red symbols) and numerically simulated (blue symbols) mean dwell time, $\langle t_d\rangle/t_0$, as a function of the effective barrier height $E_b(v_0)/k_BT$ obtained for different values of $v_0$. Here, $\langle t_d\rangle$ is normalized by the relaxation time $t_0$. The two rightmost data points (black circle and blue triangle) are obtained for passive particles with  $v_0=0$. The solid line shows an exponential fit, $\langle t_d\rangle/t_0 \propto \exp(E_b/k_BT)$, to the black circles. The error bars show the experimental uncertainties of the black circles.}
\label{Figure03}
\end{figure}

The measured $U_{\text{eff}}(x;v_0)$ has an important effect of spatial confinement on the Janus particles. From the particle trajectories, we measure the dwell time $t_d$ for a particle staying in the same groove. As shown in Fig.~\ref{Figure03}(a), the obtained PDFs $P(t_d)$ for different values of $v_0$ all collapse onto a common master curve, once $t_d$ is normalized by its mean value $\langle t_d\rangle$. The measured $P(t_d)$ has a long tail, which is well described by the exponential function, $P(t_d)\propto \exp(-t_d/\langle t_d\rangle)$ (solid line). This result suggests that the escape events occur randomly in time and can be described by a Poisson process \cite{Hanggi90,Geiseler16}.
Figure~\ref{Figure03}(b) shows how the normalized mean dwell time $\langle t_d\rangle/t_0$ changes with $E_b(v_0)$. The obtained $\langle t_d\rangle/t_0$ follows a Kramers-like relation \cite{Hanggi90,Woillez19,Reimann99}, $\langle t_d\rangle/t_0\propto \exp(E_b(v_0)/k_BT)$, for large values of $E_b(v_0)/k_BT$. This result further demonstrates that the dynamics of the slow rotating Janus particles in a trapping potential can be well described by an effective potential $U_{\text{eff}}(x;F_0)$, once $F_0$ is properly included.

To further understand the dynamics of Janus particles in a trapping potential, we conduct Brownian dynamic simulations of run-and-tumble particles (RTPs), whose motion follows the over-damped Langevin equation
\begin{equation}
\dot{{\bf r}}=v_0\boldsymbol{\hat{u}}-\mu_0 \boldsymbol{\nabla} U(\boldsymbol{r})+\sqrt{2\widetilde D_0}\boldsymbol{\eta},
\label{eq03}
\end{equation}
where $\boldsymbol{r}(t)$ is the particle position, $\mu_0=D_0/k_BT$ is the mobility of the RTPs, $\boldsymbol{\eta}$ is a Gaussian white noise with zero mean and unit variance, and $\widetilde D_0$ is an adjustable diffusion coefficient, characterizing the amplitude of the thermal noise in the system. The particle's orientation $\boldsymbol{\hat{u}}(t)$ changes stochastically to a new direction with equal probability at a Poisson rate $1/\tau_p$, which is defined via the autocorrelation function of the orientation vector $\boldsymbol{\hat{u}}(t)$:
\begin{equation}
\langle\boldsymbol{\hat{u}}(t_1) \cdot \boldsymbol{\hat{u}}(t_2)\rangle=e^{-|t_2-t_1|/\tau_p}.
\label{eq020}
\end{equation}
In our three-dimensional (3D) simulations, the external potential $U(\boldsymbol{r})$ is always 1D along the $x$-direction, i.e., $U(\boldsymbol{r})=U_0(x)$. The simulated data used in this study are obtained when the program has run over a long period with $10^5$ time steps ($10^3$ s), so that the system is in a steady-state (see supplementary Sec.~III.A for more details \cite{Note1}).

When the input parameters of the simulation, $D_0$, $\tau_p$, $U_0(x)$ and $v_0$ are chosen to be the same as those obtained from the experiment and the value of $\widetilde D_0$ is set to be the same as $D_0$, we are able to reproduce the main features of the experimental results as shown in Figs.~\ref{Figure02}-\ref{Figure03}. For example, the dashed lines in Fig.~\ref{Figure02}(a) show the simulated $U_{\text{eff}}(x;v_0)$ for different $v_0$, which are in agreement with the measured $U_{\text{eff}}(x;v_0)$ within an accuracy level of 10\%. The resulting $\Delta E_b/k_BT$ as shown in supplementary Fig.~S8(a) \cite{Note1} obeys Eq.~(\ref{eq02}) with a slightly smaller value of $a$ for $v_0\leq 0.76~\mu$m/s. The reconstructed equilibrium potentials $U_0(x)/k_BT$ from the simulated $U_{\text{eff}}(x;v_0)/k_BT$ exhibit the same scaling form, as shown in supplementary Fig.~S8(b) and Fig.~S8(c) \cite{Note1}. Furthermore, the obtained $\langle t_d\rangle$ from the simulation follows the same Kramers-like relation, as shown in Fig.~\ref{Figure03}(b). The slight difference in the absolute value of $E_b(v_0)$ between the simulation and experiment can be attributed to a small orientation bias induced by the heavier Pt-coating. As a result, the propulsion direction of the Janus particles has a slightly higher tendency to pointing upward, which gives rise to an extra net force against gravity and thus lowers the gravitational potential $U_0(x)$ (see supplementary Sec. III.C for more discussions \cite{Note1}).

The simulation also allows us to examine other effects not accessible in the experiment. First, we reduce the amplitude of thermal noise and examine its effects on $P(x;v_0)$. When the thermal noise goes to zero, we find the Janus particles start to accumulate with $P(x;v_0)$ peaked at the two force balance positions $\pm x_0$ (where $F_0$=$-\nabla U_0(x_0)\simeq kx_0$), as shown in supplementary Fig.~S9 \cite{Note1}. Similar particle accumulation near confining boundaries was also observed in previous studies \cite{Takatori16,Shen19,Volpe11}. Because of Brownian diffusion, the peak width $\sigma$ of $P(x;v_0)$ is broadened. For Sample S1, we find $\sigma\simeq (2D_0t_0)^{1/2}\simeq$ 0.17 $\mu$m, which is about twice larger than $x_0\simeq$ 0.09 $\mu$m (for $v_0=$ 0.51 $\mu$m/s). In this case, the two peaks of $P(x;v_0)$ overlap so closely (with separation $2x_0\ll 2\sigma$) that they cannot be resolved in the experiment. Our simulation results also suggest that the Brownian broadening of $P(x;v_0)$ is further enhanced when the particle's orientation is changed from 1D to 3D (see supplementary Sec.~III.B for more details \cite{Note1}).

Second, we find the measured effective potential barrier height $E_b(v_0)$ saturates at $\sim$2.8 $k_BT$ for $v_0>0.76~\mu$m/s (see Fig.~\ref{Figure02}(b)). This saturation effect is not observed in the simulation, however, where the calculated $E_b(v_0)$ continues to decrease to zero with increasing $v_0$. Recent studies \cite{Das15,Simmchen16} showed that geometric boundaries, such as a planar wall or a step, can realign the orientation of the Pt-SiO$_2$ particles via hydrodynamic interactions, so that these particles tend to move along the edge or groove of the geometric boundaries. When the particles have a higher tendency to move along the groove ($y$-direction) with increasing $v_0$, the projection of $v_0$ along the $x$-direction ceases to grow with $v_0$, and hence some of the Janus particles remain inside the groove and give rise to a saturation in the barrier height.

In summary, we have made a new experimental design to overcome the problem of sample polydispersity and identified a unique non-equilibrium regime for the slow rotating SPPs in a strong symmetric trapping potential, in which the non-equilibrium steady-state properties of the SPPs can be described by an effective equilibrium approach. First, the probability density function (PDF) of the particle's position, $P(x;v_0)$, takes a Boltzmann-like form, $P(x;v_0)\sim \exp{[-U_{\text{eff}}(x;v_0)/(k_B T)]}$, where $U_{\text{eff}}(x;v_0)$ is an activity-dependent effective potential. This result is obtained under the strong confinement condition, under which
the self-propulsion force $F_0=v_0/\mu_0$ of the SPPs is much smaller than the critical force $F_c\sim E_{b}(0)/(\lambda/2)$ needed to take away the potential barrier, so that the particle fluxes are strongly hindered and remain negligibly small \cite{Ma15}. Second, the effective potential $U_{\text{eff}}(x;v_0)$ takes the simple form given in Eq.~(\ref{eq02}), once the effects of $F_0$ are properly included under the fixed angle approximation \cite{Pototsky12,Geiseler16}. Third, with the effective potential $U_{\text{eff}}(x;v_0)$, the barrier-crossing dynamics of the SPPs follows a Kramers-like relation \cite{Hanggi90,Woillez19,Reimann99}. Finally, in the confined space, the Brownian motion of the SPPs including both translation and 3D rotation plays an important role in determining the broadening of the PDF $P(x;v_0)$. These new findings provide a coherent guideline for future experimental studies and for the development of novel applications of the SPPs in confined geometries, such as advanced technologies in microfluidics and targeted drug delivery. Incidentally, the non-equilibrium regime discussed here is very different from the universal effective equilibrium limit, as reported in Refs.~\cite{Palacci10,Maggi14} and reviewed in Ref.~\cite{Byrne22}, where the persistence time $\tau_p$ of the SPPs is short compared to the time scale of interest. In this short $\tau_p$ limit, the work done by the self-propulsion force, $W = F_0\ell_p  \propto F_0^2$, is limited by the persistence length $\ell_p = v_0\tau_p$, and thus gives rise to an elevated effective temperature \cite{Palacci10,Maggi14,Choudhury17}.

\begin{acknowledgments}
The authors wish to thank P. Fischer, D. Ou-Yang, T.-Z. Qian and W. Wang for useful discussions. This work was supported in part by RGC of Hong Kong under grant nos. 16302718 (P.T.) and 16300421 (P.T.) and by MoST of Taiwan under the grant no. 110-2112-M-008-026-MY3 (P.Y.L.). X.X. is supported by NSFC for Young Scientists of China (No.~12004082) and by Guangdong Province Universities and Colleges Pearl River Scholar Funded Scheme (2019). Y.W. and Z.L. contributed equally to this work.
\end{acknowledgments}

\end{document}